\documentclass[prl,twocolumn,showpacs,superscriptaddress,preprintnumbers,amsmath,amssymb]{revtex4}
\usepackage{dcolumn}
\usepackage{bm}
\usepackage{graphicx}
\usepackage{bm}
\begin{document}
\title{Direct evidence of random field effect on magnetic ordering of La$_{0.5}$Gd$_{0.2}$Sr$_{0.3}$MnO$_3$ manganite system}
\author{P. Dey}
\affiliation{Department of Physics and Meteorology, Indian Institute of Technology Kharagpur,
W.B., ~721~302, INDIA}
\author{T.K. Nath}
\affiliation{Department of Physics and Meteorology, Indian Institute of Technology Kharagpur,
W.B., ~721~302, INDIA}
\author{A. Banerjee}
\affiliation{UGC-DAE Consortium for Scientific Research (CSR),\\ University Campus, Khandwa Road, Indore, 452 017, INDIA}
\date{\today}
\begin{abstract}
In this Letter, we report direct experimental evidence of the effect of quenched random field on magnetic ordering of La$_{0.5}$Gd$_{0.2}$Sr$_{0.3}$MnO$_3$ manganites system. We demonstrate magnetic measurements providing serious evidence to support that quenched random field of Gd divides the system in finite size clusters which undergo cluster-glass like freezing. The size or concentration of these clusters is found to be closely related to the effect of random field, which inturn is a function of applied field. 
\end{abstract}
\maketitle 
Quenched disorder, even in arbitrarily small concentration, is expected to modify qualitatively the critical behavior of pure systems in various situations. In a classic paper, Imry and Ma \cite{imry} argued that, for isotropic case, the ordered state of a very large system is unstable against an arbitrarily small random field (H$_R$). There are number of situations where quenched H$_R$ appear to be physically realizable such as random sources and sinks of superfluid particles, stray magnetic field, liquid crystals, displacive transitions or various kinds of electronic instabilities leading to lattice distortions where defects, impurities and dislocations may couple to the order parameter. Very recently, effect of quenched disorder has been investigated in diversified context such as directed polymer \cite{Giacomin}, liquid crystal \cite{Popa}, specific heat problem \cite{Malakis}, system having coexistence of superconducting and insulating states \cite{Littlewood} in the light of model of H$_R$ \cite{imry}. Another interesting case may be a classical ferromagnet with antiferomagnetic (AFM) site impurities; a finite applied magnetic field will have reversed sign on those impurity spins leading to random component of field  \cite{imry}. Magnetism with the effects of H$_R$ is one of the areas that receives continuous interest \cite{ fishman, wolter, Tissier}. Recently, the effect of H$_R$ is shown to be responsible for suppression of ferromagnetism by a relatively small number of sites with local negative field in dilute magnetic semiconductor \cite{Bouzerar}. Kimura {\em et al.} \cite{Kimura} showed in Nd$_{0.5}$Ca$_{0.5}$Mn$_{0.98}$Cr$_{0.02}$O$_3$ manganites, the random Cr doping induces a quenched H$_R$ leading to relaxor like ferromagnetism. Rivadulla {\em et al.} \cite{Rivadulla} propose a general mechanism in which H$_R$, introduced by the fluctuations in the magnetic/orbital ordering due to proximity to localized transition, break up the system in clusters which describe the suppression of magnetic phase transition close to metal-insulator crossover. C. Sen {\em et al.} \cite{sen} explained the insulator to metal transition in manganites occuring as a result of quenched disorder in the form of  local random energies. In a very recent report \cite{Leighton}, glassy transport phenomena in a phase separated perovskite cobaltite has been understood within the model of H$_R$ \cite{imry}. All these reports in diversified physical systems indicate immense relevance of H$_R$ effect in any real system. However, the phase behaviors and how the ordering transitions of a system are affected by the presence of a weak H$_R$ remains an unsettled problem for decades. From experimental point of view, a true realization of random field model \cite{imry} is hardly conceived. Although, it has been shown \cite{fishman} that dilute antiferromagnets in uniform external field represent physical realizations of  the effect of H$_R$, experiments on such systems have proven to be very difficult and their interpretation doubtful due to slow, glassy dynamics.  

Gadolinium (Gd) substituted La$_{0.7-x}$R$_x$Sr$_{0.3}$MnO$_3$ [R = Gd] system is another the most relevant for the study of this nature, since, several groups \cite{sun, terai, snyder, hemberger} have established from different measurements that the Gd spins are polarized antiparallel to the ferromagnetic (FM) Mn sublattice resulting in Gd sublattice being antiparallel to Mn sublattice. For low doping level of Gd, it would be justified to consider those Gd$^{3+}$ ions site as AFM impurity sites in a double-exchange FM background of Mn sublattice because of stronger 3d exchange interaction compared to that of 4f, we expect Mn-Mn intra sublattice coupling to be stronger than Mn-Gd inter sublattice coupling, and Gd-Gd intra sublattice coupling to be negligible. Thus the quenched H$_R$ may be physically realized in finite applied magnetic field with spins of Gd$^{3+}$ ions polarized in the opposite direction in the background of the FM  Mn sublattice, leading to a random component of field \cite{imry}. We provide serious experimental evidence to support a scenario in which strange collective/glassy behavior can be explained only considering the effect of H$_R$ on magnetic ordering of La$_{0.5}$Gd$_{0.2}$Sr$_{0.3}$MnO$_3$ system. In this Letter, we report an unprecedented and very direct experimental evidence of the effect of H$_R$ on magnetic ordering of manganites, which further proves the tremendous relevance of the effect of H$_R$ in case of any real system having several unusual effects. Moreover, to the best of our knowledge, the fundamental prediction \cite{weinan} of nonmonotonic dependence of domain size (L) on correlation length ($\xi_R$) of H$_R$ is experimentally realized for the first time with the predicted crossover behavior of L with $\xi_R$ in this system.     

For this purpose, polycrystalline La$_{0.5}$Gd$_{0.2}$Sr$_{0.3}$MnO$_3$ sample is prepared using solid-state ceramic route. The structure and phase purity of as-prepared samples is checked by powder x-ray diffraction. We have undertaken detailed magnetic study using linear and nonlinear ac-susceptibility ($\chi_{ac}$) measurements. Linear $\chi{_{ac}}$, measured at a fundamental frequency {\em (f)} of 131 Hz and at different ac magnetic field (H$_{ac}$), shows a broad rounded maximum, where the exact position of the shoulder as well as of its H$_{ac}$ and {\em f} dependence is rather difficult to determine. On the contrary, since nonlinear effects become more pronounced in the vicinity of phase transition, the hump/peak as well as their H${_{ac}}$ and {\em f} dependence becomes much clearer. Moreover, subtle features not discernible using linear $\chi{_{ac}}$ can be picked up using non-linear $\chi{_{ac}}$. It has already been established how nonlinear $\chi_{ac}$ can be used to unravel the magnetism of interesting metastable magnetic systems and the associated short-range order \cite{sunil, bajpai} as well as to effectively probe the critical behavior of systems with long-range magnetic order \cite{alok}. So, in this present work, we have concentrated mainly on the behavior of second ($\chi {_2}$) and third ($\chi {_3}$) order $\chi_{ac}$. $\chi {_3}$ has long been used as a direct probe of the divergence of Edward-Anderson order parameter, signifying the onset of a spin-glass (SG) transition \cite{binder}, where theoretically, $\chi {_3}$ is expected to have a negative divergence in the limits H${_{ac}}$ $\rightarrow$ 0, and T $\rightarrow$ T$_G$ [T$_G$ = SG freezing temperature] \cite{suzuki}. Generally, this is demonstrated by plotting the magnitude of the peak value of $\chi {_3}$ ($\vert \chi {_3}{^{max}}\vert$) as a function of H$_{ac}$ as shown in Fig. 1, where $\vert \chi {_3}{^{max}}\vert$ is plotted as a function of H$_{ac}$ for the sample La$_{0.5}$Gd$_{0.2}$Sr$_{0.3}$MnO$_3$, from which one can conclude that $\vert \chi {_3}{^{max}}\vert$ diverges in the limit of H$_{ac}$ $\rightarrow$ 0 indicating the magnetic phase occurs due to cooperative freezing phenomenon. This divergence is further substantiated by log-log plots of $\vert \chi {_3}{^{max}}\vert$ against H$_{ac}$ [inset (a) in Fig. 1] where the negative slope of the straight line that can be considered as associated critical field exponent, is found to be -0.85 ($\pm$ 0.03), which is in close agreement with -1.1 ($\pm$ 0.04) as estimated for another SG-like system \cite{bajpai}. Moreover, we have also shown [inset (b) in Fig.1] a log-log plot of  $\chi {_3}$ against reduced temperature (T - T$_G$)/T$_G$, showing criticality of $\chi {_3}$ as a function of temperature \cite{bajpai, binder}. The exponent associated with $\chi {_3}$ -1.2 ($\pm$ 0.05) remains close to the field exponent and that of random-bond Ising spin glass ($\nu{_3}$ = 1.18 $\pm$ 0.04). Based on this analysis we may conclusively assert that the magnetic phase associated with our system reflect SG dynamics, whether it constitutes a classical SG or the behavior is introduced by random dipolar inter cluster interactions among magnetic clusters, i.e. cluster glass (CG). 

Temperature associated with the peak in $\chi {_3}$ that shows divergence in the limit of H${_{ac}}$ $\rightarrow$ 0, is the corresponding freezing temperature (T${^{\star}}$) [ inset (a) in Fig. 2] \cite{suzuki}. Figure 2 shows unusual dependence of T${^{\star}}$ on H$_{ac}$ clearly indicating an initial sharp rise in T${^{\star}}$ with H$_{ac}$ up to a certain value and then decreases gradually with field. Most interestingly, T${^{\star}}$ also show similar crossover behavior with superimposed dc field (H$_{dc}$) at the same H$_{ac}$ [ inset (b) in Fig. 2]. To the best of our knowledge there is no report till date regarding such strange field dependence of T${^{\star}}$ of any kind of glassy system. Figure 3 and its inset (a) show $\chi{_3}$ and $\chi{_2}$ at H$_{ac}$ = 2 Oe, respectively. A distinct kink, as indicated by arrow in Fig. 3, is clearly noticed in $\chi{_3}$ at  268 K, higher than the temperature (T${^{\star}}$ = 243 K) of the peak arises as a result of cooperative freezing phenomena, as has already been established in previous paragraph. A similar kink at almost same temperature of 268 K is also visible in out-of-phase linear $\chi_{ac}$ ($\chi{_1}{^I}$) [not shown]. More crucial point is that these kinks in $\chi{_3}$ and $\chi{_1}{^I}$ are accompanied by a very distinct and sharp peak in $\chi{_2}$ at exactly the same temperature of 268 K [inset (a) in Fig. 3]. However, corresponding to the freezing temperature T${^{\star}}$ = 243 K, there is not at all any response of $\chi{_2}$. As it is well known that even order $\chi_{ac}$ ($\chi{_2}$, $\chi{_4}$..) can be observed only if the system exhibits spontaneous magnetization (M$_S$) due to lack of inversion symmetry with respect to H$_{ac}$, thus we may vividly correlate the peak in $\chi{_2}$ as well as the kinks in $\chi{_3}$ and $\chi{_1}{^I}$ at 268 K with the formation of finite size FM clusters, which subsequently undergo random dipolar intercluster interactions and finally freeze to give rise a peak at a lower temperature (T${^{\star}}$ = 243 K). Thus it reveals that the associated magnetic phase with the system is CG phase. The absence of any response of $\chi{_2}$ at the temperature position (T${^{\star}}$ = 243 K) of this lower temperature peak at $\chi{_3}$ further substantiates its glassy origin \cite{sinha}. Similar features of $\chi_{ac}$ are also seen in case of other low H$_{ac}$s [insets (a) and (b) in Fig. 3, at H$_{ac}$ = 3 Oe].

It is well established that increase in cluster concentration in a system causes enhancement of random dipolar intercluster interactions that in turn increases frustration and collectivity observed in the relaxation of the system, which shift freezing temperature towards higher temperature side \cite{jonson, dormann}. Following this, within the understanding of CG phase of our system, our primary conjecture to address the rise in T${^{\star}}$ with field [Fig. 2] is the increase in cluster concentration in the system with field; likewise the subsequent decrease in T${^{\star}}$ can be correlated with the decrease in cluster concentration. It can be well understood that modulation of cluster concentration can take place as a result of modulation of cluster size, viz., decrease in cluster size causes increase in cluster concentration in the system and vice versa. Our conjecture of increase in cluster concentration with H$_{ac}$ up to a certain limit, which is supposed to take place as a result of decrease in cluster size, can be substantiated by the corresponding decrease in absolute value of $\chi{_2}$ with increase in H$_{ac}$ (not shown). Basically, decrease in cluster size can occur as a result of decrease in magnetic correlation length $\xi$, which in turn implies reduction in M$_S$ in the system hence an expected signature is decrease in corresponding absolute value of $\chi{_2}$. But from this point one cannot conclusively assert whether this decrease in $\chi{_2}$ with increase in H$_{ac}$ is associated with decrease in M$_S$ in the system or this is just an effect of decreasing nonlinearity of the system with increase in H$_{ac}$. Thus to avoid this controversy we are interested to see the effect of H$_{dc}$ on $\chi{_2}$ at the same H$_{ac}$. Figure 4 reveals that with the superposition of H$_{dc}$ = 0.1 Oe, $\chi{_2}$ peak decreases and finally it disappears with the application of  H$_{dc}$ up to 4 Oe [inset (a) of Fig. 4]. But with the application of H$_{dc}$ = 7 Oe, the same peak again appears and further grows with the application of H$_{dc}$ = 15 Oe [inset (b) of Fig. 4]. The broad low temperature peak in $\chi{_2}$, which is absent for H$_{dc}$ = 0 Oe, can be understood in terms of biasing effect of spin-clusters on the glassy background in presence of H$_{dc}$ resulting in a strong component of symmetry breaking field inside the system. However, it should be noted that this broad low temperature peak cannot be related with spin freezing, as for spin freezing phenomena $\chi{_2}$ would be essentially absent \cite{sinha}. These experimental results now unambiguously establish the fact that in this system with the increase in applied field up to a certain limit there is a gradual destruction of long range FM ordering ($\xi$ becomes smaller) resulting in decrease in cluster size, as indicated by the corresponding suppression and finally disappearance of $\chi{_2}$ up to H$_{dc}$ $\sim$ 4 Oe [Fig. 4]. However, beyond that certain limit of field there is again an increase in $\xi$ in the system causing increase in cluster size, as indicated by the reappearance of $\chi{_2}$ [Fig. 4]. Most interestingly, the crossover field (H$_{dc}$ $\sim$ 4 Oe) where disappearance and then reappearance of $\chi{_2}$ with H$_{dc}$ occur [Fig. 4] is consistent with that H$_{dc}$ where T${^{\star}}$ also show similar crossover behavior [inset (b) of Fig. 2]. Thus the variation of  T${^{\star}}$ with H$_{dc}$, as well as H$_{ac}$ [Fig. 2 and its inset (b)], can be conclusively attributed to the modulation of cluster size with applied field, which in turn modulates cluster concentration that tune the frustration and collectivity observed in the relaxation of the system and consequently tune T${^{\star}}$.            

This typical unprecedented observation of concentration/size modulation of clusters with applied field can be understood only considering H${_R}$ effect on magnetic ordering of the respective system. It has  already been mentioned that the expected magnetic state of the system consists of a double exchange FM background of Mn sublattice with AFM impurity Gd${^{3+}}$ ion sites. An applied magnetic field will have reversed sign on those AFM Gd${^{3+}}$ ion sites thus leading to components of H${_R}$ \cite{imry}, which destroy the long range ordering and break the system into clusters. Our experimental observation of decrease in cluster size with increase in applied field up to a certain limit, followed by a gradual increase in cluster size with further increase in field indicate following probable phenomenon. At low applied fields, Mn sublattice is FM with the Gd sublattice coupled antiferromagnetically with Mn sublattice. In our case, though any kind of direct correlation is not expected among Gd$^{3+}$ spins, i.e., H${_R}$ site, but as we keep on increasing applied field, Mn sublattice will become more and more stronger, which in turn influences more and more Gd$^{3+}$ spins to align (antiparallel) in the exchange field of Mn moments; thereby increasing correlation length ($\xi{_R}$) of H${_R}$.  It appears that with increase in applied field up to a certain limit Gd sublattice gets more correlated so the correlation ($\xi{_R}$) between Gd${^{3+}}$ spins increases, enhancing the AFM H${_R}$ effect resulting in decrease in cluster size. However, beyond a certain field because of higher spin moment of Gd${^{3+}}$ (7/2 $\hbar$) they would like to align towards the field \cite{Rudolf} which may essentially cause a change in the behaviour of $\xi{_R}$, i.e., H${_R}$ effect in the system resulting in increase in the cluster size. A similar result was also theoritically predicted by Weinan E {\em et al.} \cite{weinan}, which states a distinct crossover behavior of cluster/domain size L, formed in ordered systems in presence of quenched H${_R}$, as a function of $\xi{_R}$ of H${_R}$. Although experiments on different systems are consistent with this prediction \cite{weinan} in different regimes of $\xi{_R}$, to the best of our knowledge \textit{this is the first report where the predicted crossover behavior of L with $\xi_R$ is experimentally realized.}      

In conclusion, we have shown that the quenched random field of Gd divides the La$_{0.5}$Gd$_{0.2}$Sr$_{0.3}$MnO$_3$ system in finite size clusters which undergo cluster-glass like freezing. The size or concentration of these clusters is closely related to the effect of random field, which inturn is a function of applied field. We understand that our experimental result gives a very direct evidence of the effect of random field on magnetic ordering of the system thus further establishing the immense relevance of H${_R}$ effect in any real system having several unusual effects.

\newpage
\begin{figure*}
	\centering
		\includegraphics{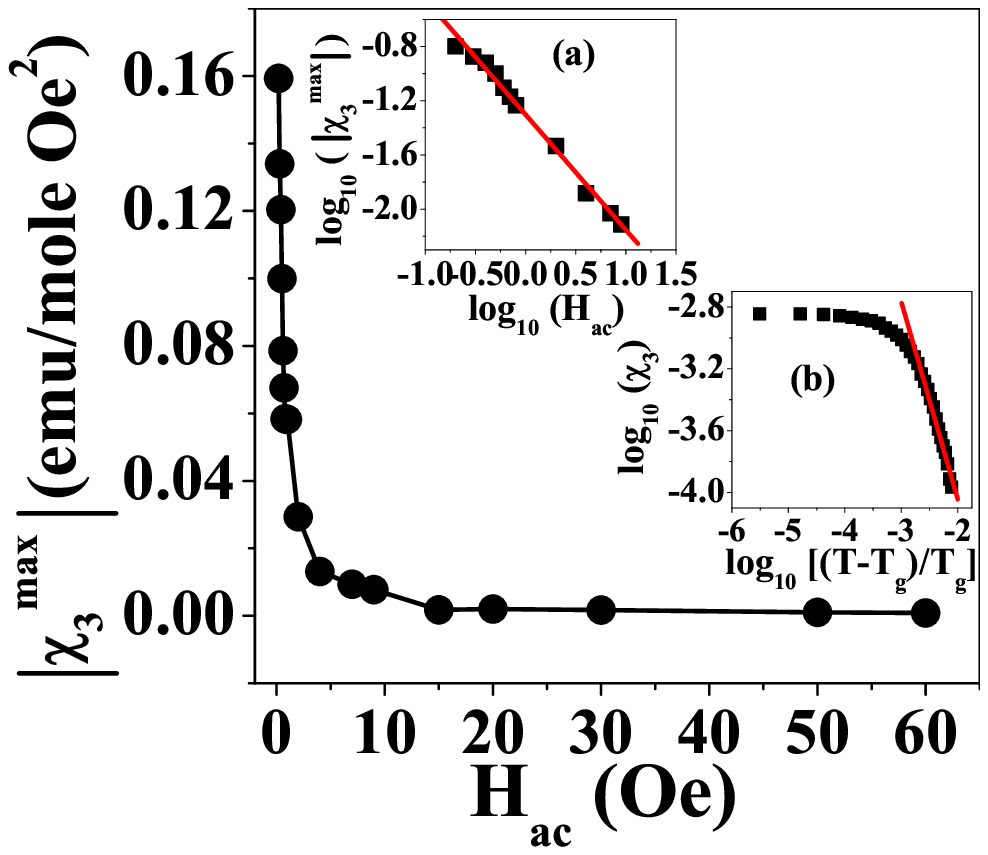}
	\caption{(Color online) $\vert \chi {_3}{^{max}}\vert$ plotted as a function of H${_{ac}}$, clearly showing divergent nature of $\chi{_3}$ at H${_{ac}}$ $\rightarrow$ 0, and T $\rightarrow$ T${_G}$, indicating SG-like magnetic phase. Insets (a) and (b) show a log-log plot of $\vert \chi {_3}{^{max}}\vert$ against H${_{ac}}$ and a log-log plot of $\chi{_3}$ against reduced temperature (T - T$_G$)/T$_G$, respectively. The straight lines are the linear fit to the data in insets.}
	\label{fig:Fig1}
\end{figure*}

\begin{figure*}
	\centering
		\includegraphics{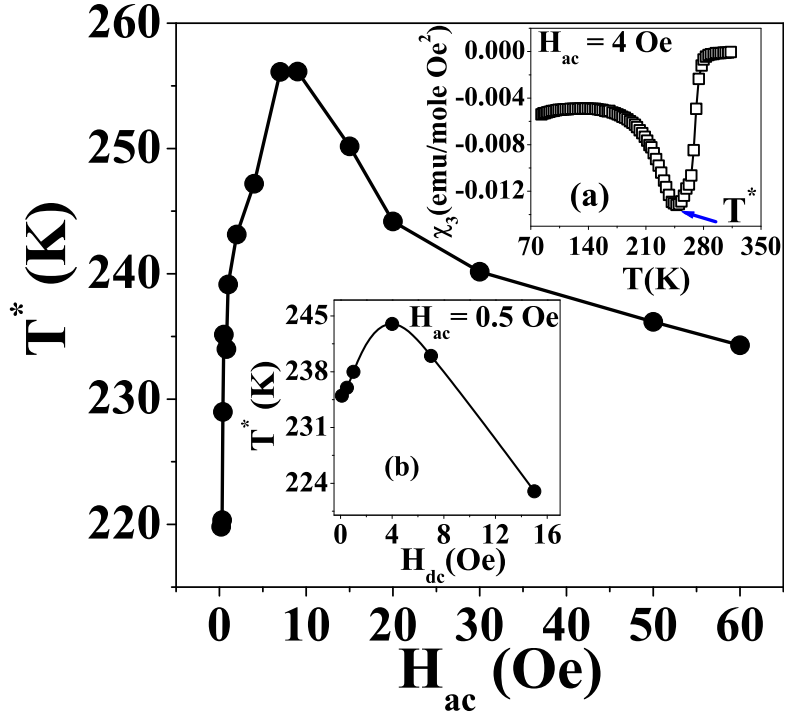}
	\caption{(Color online) H$_{ac}$ dependence of freezing temperatures T${^{\star}}$, estimated from the peak of $\chi {_3}$  versus T curves. Inset (a) shows $\chi {_3}$ versus T curve at H$_{ac}$ = 4 Oe, where the arrow indicates freezing temperature T${^{\star}}$. Inset (b) shows H$_{dc}$ dependence of T${^{\star}}$ at H$_{ac}$ = 0.5 Oe}
	\label{fig:Fig2}
\end{figure*}

\begin{figure*}
	\centering
		\includegraphics{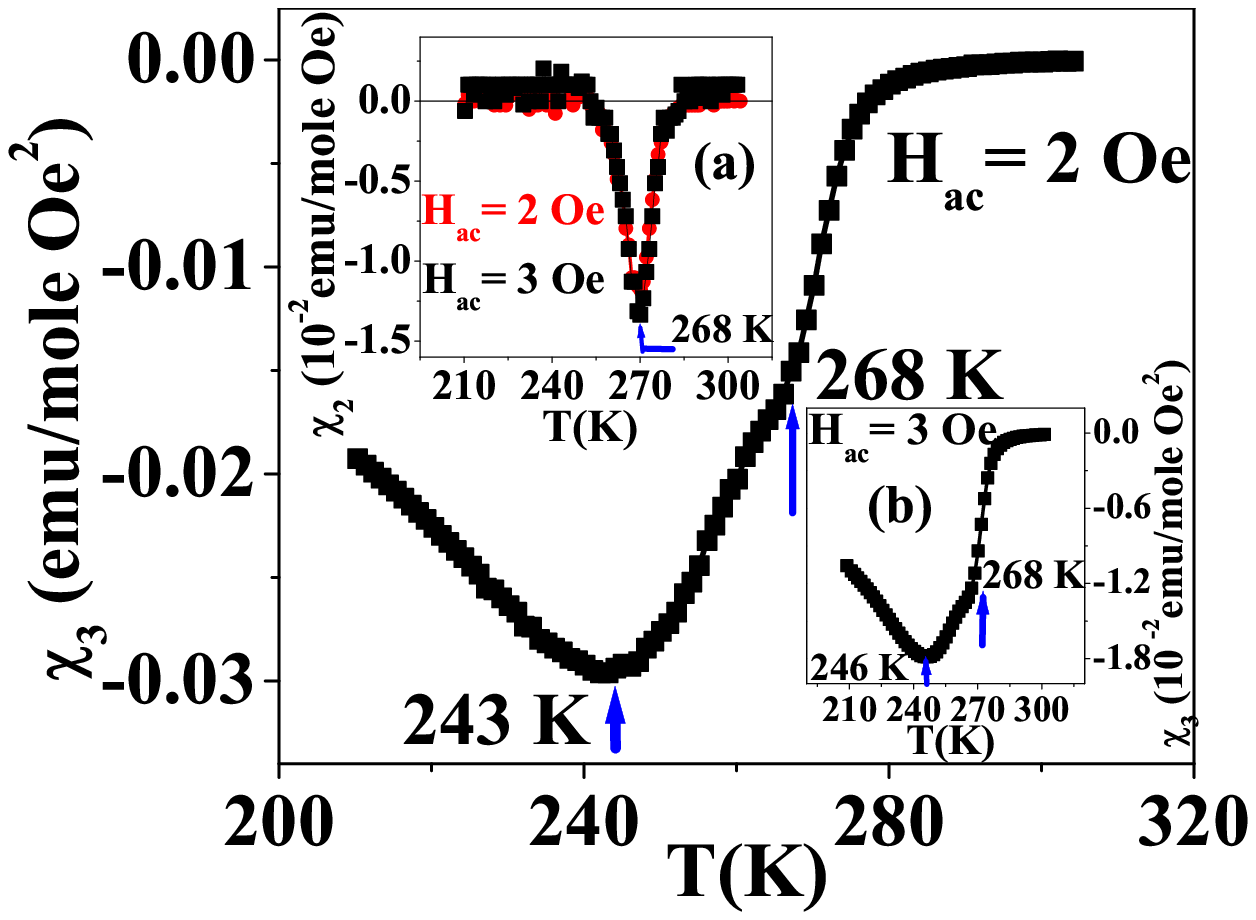}
	\caption{(Color online) $\chi{_2}$ versus T curves measured at H$_{ac}$ = 0.5 Oe, with H${_{dc}}$ = 0 Oe (unfilled circle) and 0.1 Oe (unfilled triangle). Inset (a) shows the same plot for H${_{dc}}$ = 4 Oe, whereas inset (b) shows for H${_{dc}}$ = 7 Oe (filled square) and 15 Oe (filled circle).}
	\label{fig:Fig3}
\end{figure*}

\begin{figure*}
	\centering
		\includegraphics{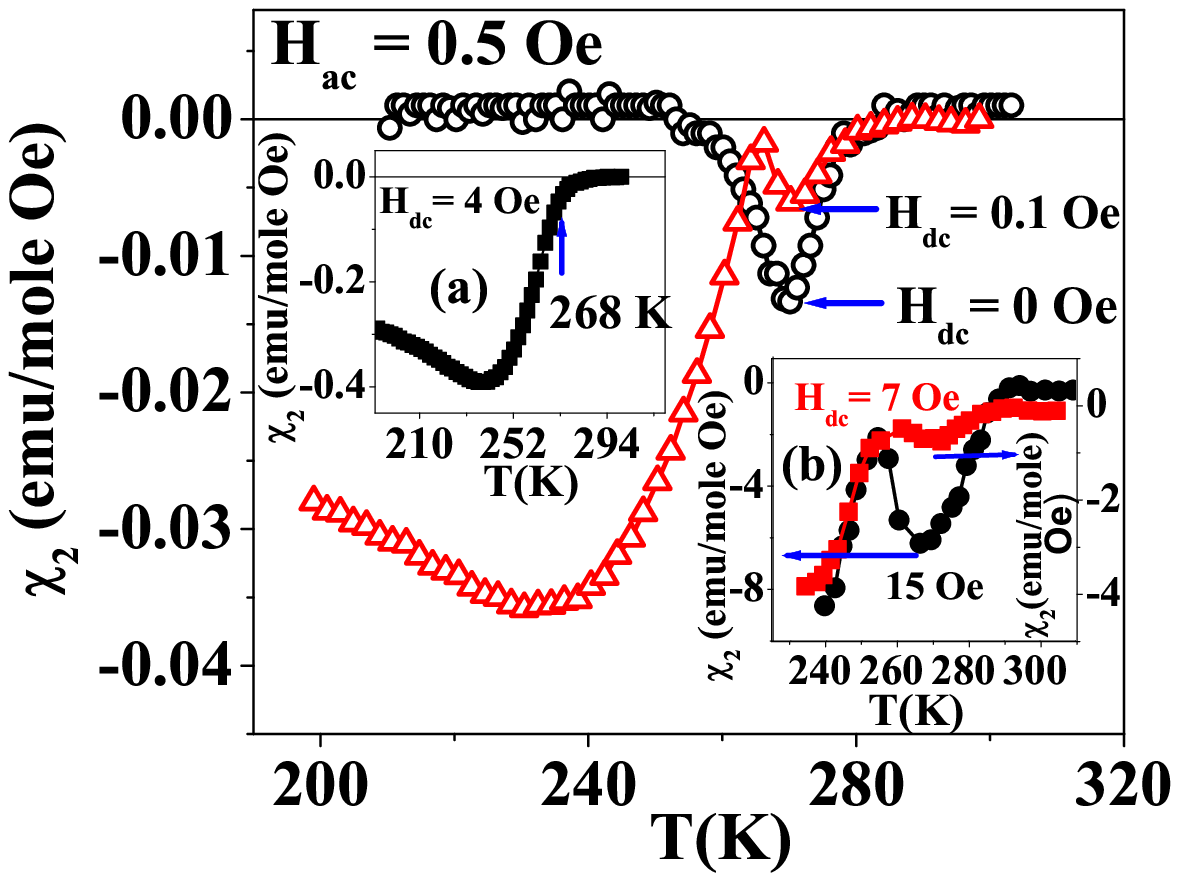}
	\caption{(Color online) $\chi{_2}$ versus T curves measured at H$_{ac}$ = 0.5 Oe, with H${_{dc}}$ = 0 Oe (unfilled circle) and 0.1 Oe (unfilled triangle). Inset (a) shows the same plot for H${_{dc}}$ = 4 Oe, whereas inset (b) shows for H${_{dc}}$ = 7 Oe (filled square) and 15 Oe (filled circle).}
	\label{fig:Fig4}
\end{figure*}

\end{document}